\documentclass[preprint]{vgtc}               

\ifpdf
  \pdfoutput=1\relax                   
  \usepackage{graphicx}                
  \DeclareGraphicsExtensions{.pdf,.png,.jpg,.jpeg} 
\else
  \usepackage{graphicx}                
  \DeclareGraphicsExtensions{.eps}     
\fi%

\graphicspath{{figures/}{pictures/}{images/}{./}} 

\usepackage{microtype}                 
\PassOptionsToPackage{warn}{textcomp}  
\usepackage{textcomp}                  
\usepackage{mathptmx}                  
\usepackage{times}                     
\usepackage{cite}                      
\usepackage{tabu}                      
\usepackage{booktabs}                  
\usepackage{amsmath}

\onlineid{0}

\vgtccategory{Research}

\vgtcinsertpkg

\title{Bring Your Own Character: A Holistic Solution for\\Automatic Facial Animation Generation of Customized Characters}

\author{Zechen Bai${^{1,2}}$\thanks{Project initiated at ISCAS. The author now affiliated with NUS. }  %
\quad Peng Chen${^{1,3}}$  %
\quad Xiaolan Peng${^{1}}$  %
\quad Lu Liu${^{1}}$ %
\quad Hui Chen${^{1,3}}$\thanks{Corresponding author} %
\quad Mike Zheng Shou${^{2\dagger}}$  %
\quad Feng Tian${^{1,3}}$  %
\\
\scriptsize{Institute of Software, Chinese Academy of Sciences, China$^1$} \\
\scriptsize{Show Lab, National University of Singapore, Singapore$^2$} \\
\scriptsize{University of Chinese Academy of Sciences, China$^3$}
}

\abstract{
Animating virtual characters has always been a fundamental research problem in virtual reality (VR).
Facial animations play a crucial role as they effectively convey emotions and attitudes of virtual humans.
However, creating such facial animations can be challenging, as current methods often involve utilization of expensive motion capture devices or significant investments of time and effort from human animators in tuning animation parameters.
In this paper, we propose a holistic solution to automatically animate virtual human faces. 
In our solution, a deep learning model was first trained to retarget the facial expression from input face images to virtual human faces by estimating the blendshape coefficients. This method offers the flexibility of generating animations with characters of different appearances and blendshape topologies.
Second, a practical toolkit was developed using Unity 3D, making it compatible with the most popular VR applications. The toolkit accepts both image and video as input to animate the target virtual human faces and enables users to manipulate the animation results.
Furthermore, inspired by the spirit of Human-in-the-loop (HITL), we leveraged user feedback to further improve the performance of the model and toolkit, thereby increasing the customization properties to suit user preferences.
The whole solution, for which we will make the code public, has the potential to accelerate the generation of facial animations for use in VR applications.
\url{https://github.com/showlab/BYOC}
} 

\CCScatlist{
  \CCScatTwelve{Virtual Human}{Facial Animation}{Blendshape}{Human-in-the-loop}
}

\begin{document}

\firstsection{Introduction}

\maketitle

Virtual reality (VR) technology is attracting increasing attention with the popularity of Metaverse in recent years.
When building virtual characters, it is important to create vibrant and vivid facial animations, as they can intuitively convey emotion and feeling, playing a crucial role in a wide range of VR applications \cite{badler1999animation,machidon2018virtual,burden2019virtual}, such as enhancing emotionally challenging experiences in VR games \cite{peng2019beyond,peng2020palette,peng2022detecting,peng2023challengedetect}.
However, it is non-trivial to create diverse facial animations conveniently.
Traditional methods usually use dedicated motion capture devices to track the key points of human actors and replicate them on virtual characters.
Approaches in industry often involve creating facial animations by combining muscle movements using coefficients. In this method, a facial expression is typically regarded as a superposition of the movement of different muscles in the face. 
As a consequence, a practical and efficient face representation called blendshape has been devised for controlling the facial expressions of avatars.
Each blendshape represents a modeled single action unit of the face with respect to the neutral face.

\begin{figure}
    \centering
    \includegraphics[width=0.99\linewidth]{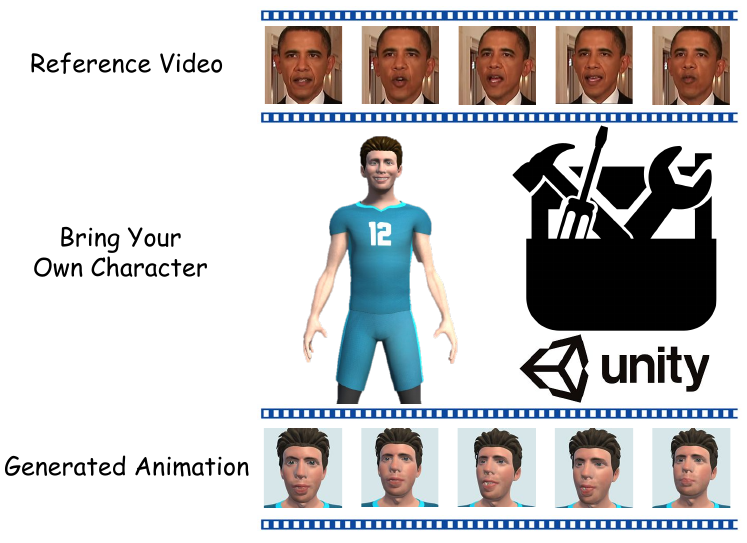}
    \caption{Given a target facial video as reference, bring your own character into our solution integrated with Unity3D, and it automatically generates facial animation for the virtual character.}
    \vspace{-10pt}
    \label{fig:teaser}
\end{figure}

Currently, facial blendshape coefficients are mainly obtained from human expert.
Through a deep understanding of the meaning of each blendshape, human animators often need to determine blendshape coefficients based on their imagination of the desired facial expression. Alternatively, they may fine-tune the coefficients using a trial-and-error approach, referencing a desired facial expression.
Both ways are time-consuming and labor-intensive.
Recently, some commercial softwares~\cite{arkit,livelink} provide the function of computing blendshape coefficients based on video or audio input.
They usually project the input signal to a pre-defined blendshape space, showing promising result.
However, in the wider range of VR applications, it is common to introduce customized virtual characters holding appearances and topologies that are different from the pre-defined ones.
To our knowledge, there is still a gap in the available solutions for creating animations for customized virtual characters. 
To address this, our paper proposes a comprehensive solution that automates the generation of facial animations specifically tailored to customized virtual characters with varying appearances and blendshape topologies.
By doing so, our approach aims to alleviate the workload of human animators and expand the scope of use and application.
The main idea of our solution is illustrated in Fig.~\ref{fig:teaser}.

We first present a deep learning model to retarget the reference facial expression image.
This is achieved by estimating the blendshape coefficients that are able to replicate the reference facial expression on the target virtual character. 
The model consists of a base model and an adapter model.
The base model is responsible for extracting generic 3D facial parameters from the given reference image.
The adapter model aims to adapt the generic 3D facial parameters to specific blendshape coefficients.
Specially, the base model is character-agnostic, \textit{i.e,} once pre-trained, it is applicable to all the characters.
The adapter model is topology-aware, \textit{i.e,} it is sensitive to the blendshape topology of the virtual character.
Once trained, the whole model is compatible to virtual characters with different texture appearances and the model can also be quickly adapted to a virtual character with a new blendshape topology by simply finetuning the adapter on a automatically generated dataset.

We then develop a toolkit in Unity3D to provide a user-friendly interface to the proposed model with additional design on incorporating user feedback.
The toolkit accepts video or image as input.
For video input, the toolkit will output a virtual human animation synchronized with the video.
For image input, the toolkit generates a smooth facial animation by treating the facial expression in the image as the peak intensity.
The toolkit provides user interface to modify the generated results, such as tuning some unsatisfied blendshape coefficient values or highlighting the keyframe-of-interest.
Besides, the human feedback will be utilized to further enhance the performance of the toolkit and the internal model.
Instant feedback on certain keyframes acts as user preference for the current animation, which can be quickly applied to all the frames, accelerating the process of animation generation.
Finally, the effectiveness of the proposed deep learning method for facial animation has been evaluated, and the developed toolkit has also undergone testing to assess its functional usability and user experience.

This work contributes a holistic solution for automatic facial animation generation on customized characters, which consists of three parts:
\textbf{(1)} We present a deep learning-based method to retarget the reference facial expression by estimating blendshape coefficients. This method differs from previous works in the flexibility of generating animations with characters of different appearances and blendshape topologies, broadening the scope of use.
\textbf{(2)} We develop a toolkit that encapsulates the proposed deep learning method. The toolkit provides a practical user interface for adjustment on the fly and also leverages human-in-the-loop to enhance the performance of both the toolkit and the internal model.
\textbf{(3)} We evaluate the effectiveness of the proposed model and the usability of the toolkit, which can potentially inspire future work in the domain of virtual character animations. Additionally, we will make the entire solution's code publicly available to benefit a wider range of use cases.

\section{Related work}

\subsection{Blendshape-based Animation Creation}
Animation creation in VR has been a topic of interest in both academia and industry recently \cite{bai2021enhancing,bai2021play,bai2023simple}.
One traditional approach is the use of motion capture devices \cite{nogueira2011motion,sharma2019use}, allowing for capturing and replicating real-world movements onto virtual characters.
Another line of research focuses on parameter tuning.
By adjusting specific animation parameters, such as facial muscle movements or deformations, researchers aim to achieve expressive character animations.
Blendshapes, also known as morph targets or shape keys, is a commonly used technique in computer animation for modeling complex and subtle facial expressions and movements \cite{joshi2006learning,lewis2010direct,lewis2014practice}.
For 3D virtual human models, each blendshape represents a single facial unit, such as eyebrows, lips, jaw, etc.
By creating a series of blendshapes that represent different positions and shapes of the face, animators can create a wide range of vivid facial expressions by blending these shapes together in various combinations.
Intuitively, more blendshapes imply more detailed depiction and control of facial expression, increasing the upper bound of expressiveness.
Currently, mainstream software, such as MAYA, Blender, and Unity3D, can be utilized for editing and designing virtual humans based on blendshapes.
In such software, blendshape is implemented as the offset vectors on subset vertices of the mesh.

Recently, there are some attempts on automatically estimating blendshape coefficients from various sources, including image (or video)~\cite{arkit,volonte2022headbox,livelink}, audio~\cite{pan2023emotional}, etc.
HeadBox~\cite{volonte2022headbox} provides a facial blendshape animation toolkit specifically for the Microsoft Rocketbox Library, which supports driving the virtual character animation from real-time video input.
Reference~\cite{pan2023emotional} proposed a method that estimates the emotional expression facial parameters of virtual characters based on input audio.
JALI~\cite{edwards2016jali} is a system dedicated designed for expressive lip-synchronization animation.
Some commercial tools have been developed to support blendshape-based animation creation.
Notably, the facial retargeting product like ARKit~\cite{arkit}, Faceware suite~\cite{faceware}, and ``Live Link Face"~\cite{livelink} have gained popularity in the industry for their ability to compute blendshape weights and facilitate the replication of facial expressions on avatars.

We argue that previous works suffer from the following drawbacks: 1) they are often coupled with specific character rigs and blendshape topology, prohibiting applications on customized virtual characters; 2) the commercial software are usually expensive to use.
Our work differ from them in our support of customized virtual characters created by the users.
We will open-source the solution to benefit the community.

\subsection{Character Auto Creation}
Automatic character generation technology is based on 2D images to drive the generation of 3D models, involving estimating specific face attributes parameters from the given image.
Wolf et al.~\cite{wolf2017unsupervised} first proposed a method called Tied Output Synthesis (TOS) in 2017, which utilizes adversarial training to create parameterized avatars by selecting facial attributes from a pre-defined library of facial attribute templates.
In 2019, Shi et al.~\cite{shi2019face} proposed a method called Face-to-Parameter (F2P) that estimates a set of continuous face parameter values based on 2D facial images.
The face parameter estimation is carried out through an iterative search process.
Subsequently, Shi et al. proposed an upgraded version of the method, Face-to-Parameter V2 (F2Pv2)~\cite{shi2020fast} in their 2020 study.
This method requires only one forward neural network operation to complete the face parameter estimation, greatly improving efficiency.
The task of automatic character generation appears to be similar to blendshape coefficients estimation.
The similarity lies in that both of the problem aim to manipulate 3D characters that closely resemble the input facial images by estimating facial parameters.
However, character generation assumes that the character has a large parameter space, including both facial movements and attributes. 
In that case, the estimated parameters include identity, expression, pose, and even texture and hairstyle.
In contract, in animation creation, people mainly focus on the facial muscle movements that affect the expressions or poses, other than identity or hairstyle.
Our work lie in the line of animation creation.
This actually poses a unique technical challenge that we need to decouple and extract the expression and pose information from the facial images.

\subsection{3DMM and 3D Face Reconstruction}
3D face modeling and reconstruction is a classic problem in computer graphics, aiming at recovering 3D facial information from 2D facial images.
One representative work in this field is the 3D Morphable Model (3DMM) proposed by Blanz et al.\cite{blanz1999morphable} in 1999.
This work uses Principal Component Analysis (PCA) to parameterize the 3D mesh of the face, then optimizes the parameters of PCA to fit the input 2D facial image.
Building upon this work, new research has emerged that aims to enhance the representation power of Morphable Models in terms of shape, texture, and expression, such as BFM09~\cite{bfm09}, FaceWarehouse~\cite{cao2013facewarehouse}, and FLAME~\cite{FLAME:SiggraphAsia2017}.
3D face reconstruction based on 3DMM can be essentially regarded as a parameter fitting problem between 2D facial images and 3D facial models.
In deep learning era, this problem is extensively studied by designing a neural network model to predict 3D facial parameters based on 2D images \cite{dou2017end,feng2018joint,deng2019accurate,chaudhuri2020personalized}.
In this process, differentiable renderer is employed to render 3D face based on the predicted 3DMM parameters, then generate supervision signal to train the neural network.
This series of methods show promising performance in 3D face reconstruction but still cannot address the blendshape estimation problem.
The main problem is that differentiable renderer used in the these methods is specifically designed to fit 3DMM parameters.
It is often not available for customized virtual characters in production.
Fortunately, the universality and reliability of 3DMM have been widely verified.
We utilize 3DMM parameters of BFM09 as a parametric representation for 2D facial images in 3D space. Moreover, this 3DMM representation facilitates the explicit decoupling and extraction of expression and pose parameters.

\section{Method}
\label{sec:method}

\begin{figure*}[h]
\vspace{-10pt}
    \centering
    \includegraphics[width=0.88\textwidth]{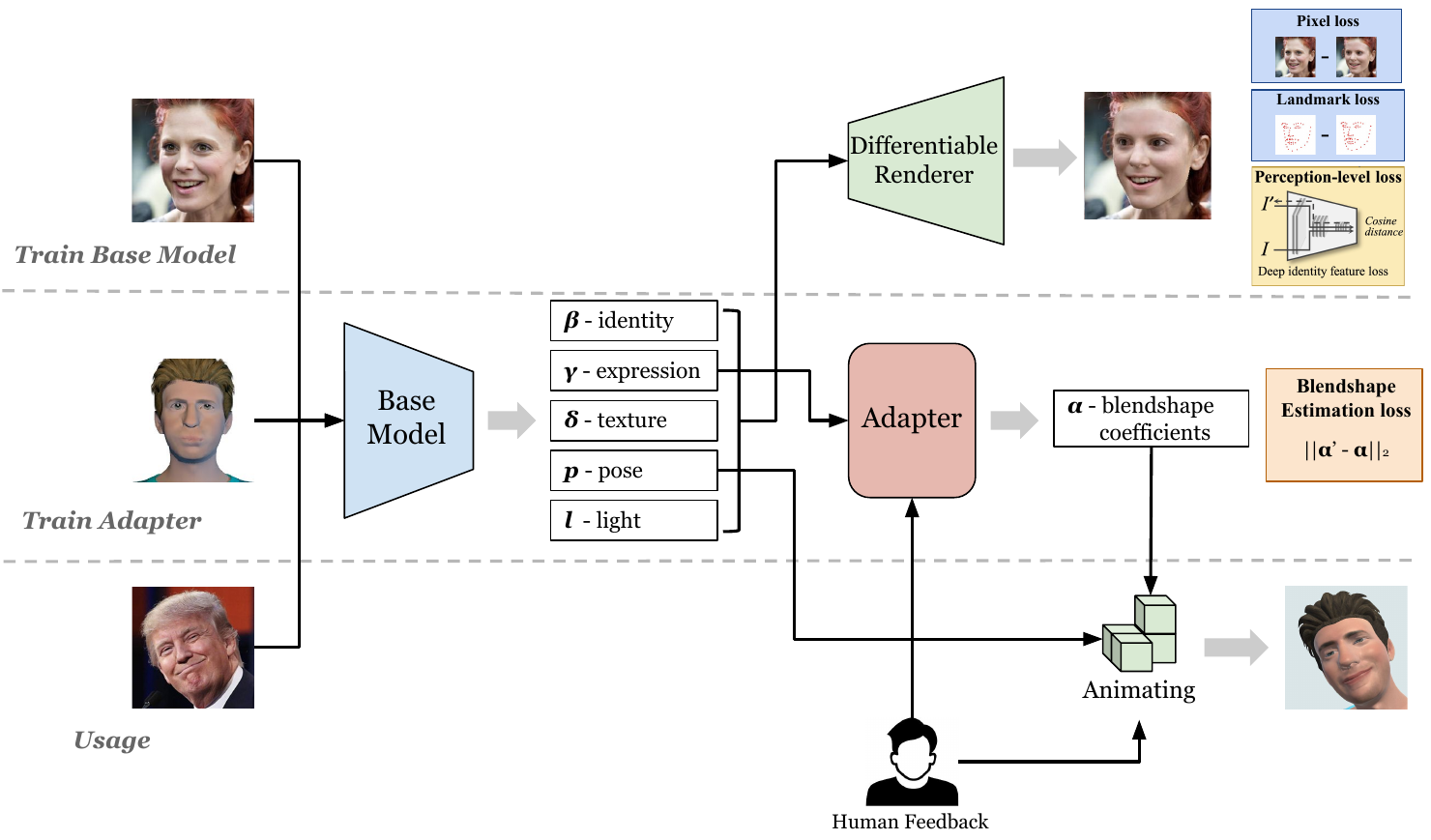}
    \vspace{-8pt}
    \caption{Illustration of training and usage of the model.
    The first stage trains the base model that regresses 3DMM facial parameters.
    The second stage trains the adapter model to estimate the target blendshape coefficients with the help of pre-trained base model.
    The usage stage utilizes the pre-trained base model and adapter model to drive virtual human to replicate facial expressions and head pose of the reference image.}
    \label{fig:method}
    \vspace{-10pt}
\end{figure*}

\subsection{Overview}
In this section, we present a deep learning-based method for estimating blendshape coefficients based on 2D facial images.
The method overview is illustrated in Fig.~\ref{fig:method}.
The task can be formally described as follows: given a 2D facial image representing a reference facial expression, the method estimates a set of blendshape coefficients that can reproduce the reference facial expression on a 3D virtual human model.
The proposed method mainly consists of two models.
First, a base model is employed to regress a set of 3DMM facial parameters based on the given facial image.
This 3DMM facial parameters can be regarded as the representation vector of a face in 3D space.
Next, an adapter model is proposed to adapt the expression parameters of 3DMM to the target blendshape coefficients.
We first introduce each individual model in Sec.~\ref{sec:base_model} and Sec.~\ref{sec:adapter} respectively.
Then we present the training and testing details in Sec.~\ref{sec:training_infer}.

\subsection{Base Model}
\label{sec:base_model}
The 3DMM parameters include identity, expression, and texture, as well as camera parameters such as pose and lighting.
Within the parameter space of 3DMM, the shape $\mathbf{S}$ and texture $\mathbf{T}$ of a face can be represented using the following approach:
\begin{equation}
    \mathbf{S} = \mathbf{S}(\beta,\gamma) = \overline{\mathbf{S}}+\mathbf{B}_{id}\beta+\mathbf{B}_{exp}\gamma,
\end{equation}
\begin{equation}
    \mathbf{T} = \mathbf{T}(\delta) = \overline{\mathbf{T}}+\mathbf{B}_{t}\delta,
\end{equation}
where $\overline{\mathbf{S}}$ represents the average face shape, $\overline{\mathbf{T}}$ represents the average face texture, $\mathbf{B}_{id}$, $\mathbf{B}_{exp}$, and $\mathbf{B}_{t}$ represent the PCA basis vectors for identity, expression, and texture, respectively.
These basis vectors is a set of orthogonal vectors obtained through principal component analysis.
Correspondingly, $\beta$, $\gamma$, and $\delta$ represent the parameters for identity, expression, and texture, respectively.
In our implementation, the popular 2009
Basel Face Model~\cite{paysan20093d} is utilized to construct $\overline{\mathbf{S}}$, $\mathbf{B}_{id}$, $\overline{\mathbf{T}}$, and $\mathbf{B}_{t}$, while the expression basis $\mathbf{B}_{exp}$ is derived from the FaceWarehouse~\cite{cao2013facewarehouse} dataset.

We employ a neural network model based on ResNet-50 for regressing the 3DMM parameters.
During training, given a RGB image $I$, the neural network model is utilized to regress the 3DMM parameters and camera parameters.
Subsequently, these parameters are employed for differentiable 3D face reconstruction and rendering, resulting in the reconstructed image $I^{'}$.
Formally,
\begin{equation}
    \beta, \gamma, \delta, \textit{p}, \textit{l}= \text{ResNet}(I),
\end{equation}
\begin{equation}
    I^{'} = \text{DifferentiableRendering}(\beta, \gamma, \delta, \textit{p}, \textit{l}),
\end{equation}
where $\beta$, $\gamma$, $\delta$, $\textit{p}$, $\textit{l}$ represent the parameters of identity, expression, texture, pose, and light illumination respectively.

Based on $I$ and $I^{'}$, this method employs a joint training approach using both image-level loss and perceptual-level loss to optimize the model.
The image-level loss comprises pixel loss and landmark loss.
The formulation of the pixel loss function is expressed as follows:
\begin{equation}
    L_{\text{photo}}(I, I') = ||I'-I||_{2},
\end{equation}
which measures the per-pixel differences between the original input image and the reconstructed image.
The landmark loss function involves detecting facial landmark $\{q_n\}$ for the original and reconstructed images using a commonly used landmark detection model \cite{king2009dlib}.
The loss function is computed as follows:
\begin{equation}
    L_{\text{lan}}(I, I') = \frac{1}{N}\sum_{n=1}^N ||q'_{n}-q_{n}||_{2},
\end{equation}
where $N$ represents the total number of landmark, which is 68 in our implementation.

In addition to employing intuitive image-level loss functions, we also incorporate perceptual loss functions.
Currently, popular perceptual loss functions typically involve utilizing pre-trained deep neural networks to extract image features and measure the differences between these features.
Considering that the algorithm in this section primarily focuses on faces, a large-scale pre-trained face recognition network, FaceNet~\cite{schroff2015facenet}, is utilized to extract facial image features.
The perceptual is as follows:
\begin{equation}
    L_{\text{per}}(I, I')=1-\frac{<f(I),f(I')>}{||f(I)|| \cdot ||f(I')|| }
\end{equation}
where $f(\cdot)$ represents the function (neural network) for extracting image features, and $<\cdot,\cdot>$ represents vector dot product. 

The base model projects the human face from a 2D image to a 3D space.
As a generic representation, the 3DMM parameters have the capacity to reproduce a large amount of human facial expressions.
Furthermore, the 3DMM parameters explicitly decouple the face into identity, expression, and pose, thereby addressing the issue of entangled facial attributes.
The base model is a character-agnostic model in this method.
Once pre-trained, it does not require any retraining when transferring the method to other 3D virtual human models with new topological structures.
In other words, this model is totally universal and reusable.

\subsection{Adapter}
\label{sec:adapter}
As shown in stage 2 of Fig.~\ref{fig:method}, the task of the adapter model takes the expression parameters as input and predicts the blendshape coefficients.
This model is a lightweight neural network model consisting of only two linear fully connected layers with an activation function between them.
After the final layer, a $Clamp$ operator is applied to truncate the output values within the range of 0 to 1, which is the valid value range of blendshape coefficients.
During the training process, mean squared error (MSE) is utilized as the loss function to optimize this model.
Given the 3DMM expression parameters $\gamma$ and the ground-truth blendshape coefficient labels $\alpha$ , this stage can be formally described as:
\begin{equation}
    \alpha^{'} = \text{Adapter}(\gamma),
\end{equation}
\begin{equation}
    \mathbf{L} = ||\alpha'-\alpha||_2,
\end{equation}
where $\alpha^{'}$ represents the output blendshape coefficients predicted by the adapter model.
The objective of model training is to minimize the loss function $\mathbf{L}$.

The adapter model is a topology-aware model.
Its training depends on specific blendshape topology of the virtual human.
In practical applications, it's inevitable to encounter virtual human models that posses different appearances, different topologies.
However, the model's generalization ability is not hindered.
Once trained, this model can be used on the virtual characters family with the same blendshape topology, even their texture appearances are different.
Besides, the model can be easily and quickly adapted to virtual characters with new blendshape topology, thanks to (1) the automatic data generation and finetuning (will be introduced in Sec.~\ref{sec:data_syn}), (2) the model's simple and lightweight architecture.
This adaptation imposes no limitation on blendshape topology. This includes, but is not limited to, the order, quantity, content, and names of blendshapes. In other words, users can provide their home-baked virtual character models with arbitrary blendshape topologies, and our method and toolkit can be utilized on their customized characters.

\subsection{Training and Testing Details}
\label{sec:training_infer}
Fig.~\ref{fig:method} shows the three stages in our method.
The first two stages introduce the training of the base model and adapter model respectively.
The third stage shows the usage of the models after training.

\subsubsection{Training Base Model}
The upper part of Fig.~\ref{fig:method} illustrates the training of the base model. We use publicly available real-world face datasets, including CelebA~\cite{liu2018large} and LFW~\cite{huang2008labeled}, to train the base model.
CelebA and LFW are high-quality face datasets with diverse and distinct variations in factors such as race, age, gender, expression, and face shape.
In the data preparation stage, we first perform face detection, alignment, and cropping using the method provided in Dlib~\cite{king2009dlib}.
Then, all images are uniformly resized to the resolution of $224\times 224$.
We follow the training scheduler of the related work~\cite{deng2019accurate}.
In short, we take the ImageNet~\cite{deng2009imagenet} pre-trained weights as the initialization, and train the base model using Adam optimizer with batch size of 5, initial learning rate of $1e-4$, and 500K total iterations.

\subsubsection{Training Adapter Model}
\label{sec:data_syn}

The middle part of Fig.~\ref{fig:method} shows the training of the adapter model.
As we mentioned, the adapter model is a topology-aware model.
Therefore, when training the adapter model on specific virtual character, or transferring to virtual characters with new blendshape topology, we propose an automatic dataset generation procedure based on the target virtual character, \textit{i.e,} character-dependent dataset.

The required dataset for a target 3D virtual character model includes: (1) randomly generated blendshape coefficients, (2) virtual human facial images rendered based on the blendshape coefficients.
Regarding the generation of blendshape coefficients, each blendshape can be treated as a channel with values ranging from 0 to 1.
In that case, the task of coefficient generation is to randomly generate a value for each channel.
For the generation of virtual human facial images, the rendering function in the 3D software, such as MAYA, Unity3D, is utilized to render virtual human face images based on the blendshape coefficients.
In our implementation, a front-facing virtual camera is used to render the virtual human face images.

For each virtual character (family) with a specific blendshape topology, we constructed a dataset of 10K data samples (8K for training, 1K for validation, and 1K for testing).
The adapter model is trained with the generated dataset.
Specifically, the rendered virtual human face images are fed into the pre-trained base model to produce 3DMM parameters.
Then, the adapter model takes the expression parameters from the 3DMM parameters as input and predicts the blendshape coefficients.
The blendshape coefficients within the dataset are used as ground-truth to supervise the predicted values.
In this process, the parameters of the base model are no longer trained, only the parameters of the adapter model are updated.

In our empirical study, we found that when generating the character-dependent dataset, injecting human prior knowledge would help stabilize the training.
Specifically, without any constraint, the blendshape coefficients are totally random values ranging from 0 to 1, which may cause invalid facial expressions.
For example, it is unlikely for a person to move their lips in two opposite directions simultaneously.
Considering this, we propose an optional step to increase the quality of the generated dataset.
This can be achieved by defining a set of rules that exclude the invalid value combinations.
With the presence of such rule-based constraints, a set of reasonable blendshape coefficient values can be obtained.

\begin{figure*}[h]
    \centering
    \includegraphics[width=0.9\textwidth]{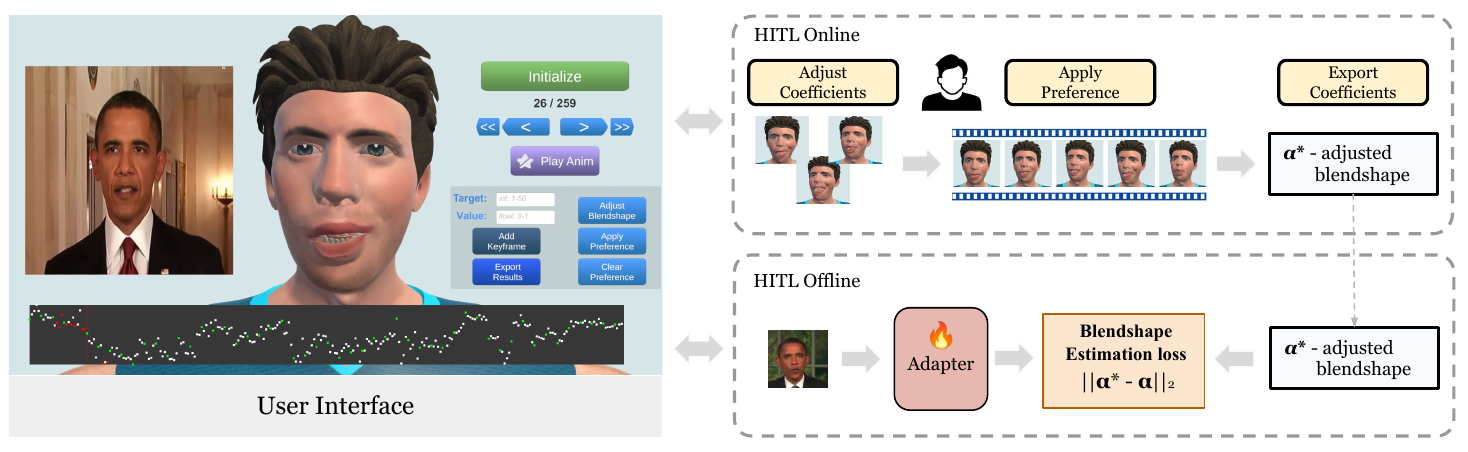}
    \vspace{-8pt}
    \caption{The left part is the user interface of the proposed toolkit. The right part is the two modes of Human-in-the-loop (HITL). In the online mode, users can adjust blendshape coefficients of several key-frames and then apply the preference to the whole video. In the offline mode, the adjusted blendshape will be collected as ground-truth data to finetune the adapter model offline, further boosting the performance of the model.}
    \label{fig:user_interface_design}
    \vspace{-12pt}
\end{figure*}

\subsubsection{Usage}
The lower part of Fig.~\ref{fig:method} shows the procedure of using the whole model.
The pre-trained base model and adapter model can be used as off-the-shelf black boxes in this phase.
Given the reference facial expression image, we first input it to the pre-trained base model to extract 3DMM parameters.
The expression parameter is used to further predict the blendshape coefficients using the pre-trained adapter model.
The estimated blendshape coefficients are used to animate the virtual character.
The pose parameter from the base model is also utilized in the animating process to increase fidelity.

In the animation generation process, apart from the two frozen models, we also introduce human feedback signals to enhance the animation quality as well as further boost the performance of the mode.
The details will be introduced in Sec.~\ref{sec:toolkit}.

\section{Toolkit}
\label{sec:toolkit}
\subsection{Holistic Design}

We develop a Unity-based toolkit that automates the process of virtual human facial animation generation based on 2D images or videos.
This toolkit extracts information of facial expression and head pose from video frames by base model, and predicts blendshape coefficients for customized virtual characters by adapter model.
The toolkit simultaneously displays the generated facial animation of the virtual character and the original video frames in the user interface, allowing for comparison and adjustment for users.
This toolkit implements a human-in-the-loop approach, where human feedback are involved into the animation generation process.
It not only injects user preference for the current video at hand, but also boost the performance of the internal deep learning model.

\subsection{Basic Functionalities}
\subsubsection{Auto-Animation Generation}
\label{sec:tool_auto_gene}
We utilize the deep learning model described in Sec.~\ref{sec:method} in the toolkit to automatically generate facial animations.
The specific generation process depends on the type of input data.
For single image input, the toolkit first estimates the blendshape coefficients based on the input image, then generates facial expression for the virtual human.
The generated expression is used as the peak intensity values.
After that, a linear interpolation is performed between the natural expression blendshape coefficient values (zeros) and the peak intensity values, gradually transitioning from zero to the peak intensity, and then back to zero.
This process ensures a smooth and gradual transitional animation that presents the input facial expression.

For video input, the toolkit first estimates the blendshape coefficients of each frame in the video.
It then sparsely samples keyframes and generates the facial animation for the virtual human by linearly interpolating between these keyframes.
The rationale of sampling with interval is that the blendshape coefficients of continuous video frames inevitably suffer from jitter problem.
This sample-and-transition strategy ensures a smooth animation between the selected keyframes.

\subsubsection{User interaction}
\label{sec:user_interaction}

\paragraph{Frame-level interaction}
When the input is a video, the toolkit adopts an sparse sampling strategy during initialization.
By default, we select one keyframe every 5 frames and generate animation.
Additionally, we provide the option for users to manually select ideal keyframes.
For example, the user may want to highlight the most exaggerate facial expression, then the neighbor frames can be all selected.
Then, the toolkit performs segmented linear interpolation between all the keyframes, including the initialized keyframes and the user-selected keyframes, to achieve a smooth animation.

\paragraph{Blendshape-level interaction}
In the toolkit, the blendshape coefficients are automatically estimated by the deep learning model.
Apart from that, we also provide the interface to let users adjust the blendshape coefficients.
For example, when the user finds that the intensity of `eye-open' blendshape does not reflect the input facial expression, he/she can easily tune the value according to his/her need.
This design offer the flexibility to correct the error produced by the model and enhance the quality of the animations.

The above modifications will be displayed in the scene in real-time for reference.
Both frame-level and blendshape-level interactions are reflection of user preference.

\subsection{Human-in-the-loop}
Human-in-the-loop (HITL) refers to a computational framework or system where human involvement is integrated into the decision-making or execution process.
It involves a collaboration between humans and machines, where humans provide input, feedback, or supervision to improve the performance of automated algorithms or systems.
Apart from automatic generation, our toolkit places emphasis on human feedback and preferences, incorporating human adjustment of blendshape parameters as a crucial element in the HITL process, boosting performance beyond model from Sec.~\ref{sec:method}. The HITL in the toolkit has online mode and offline mode.

\paragraph{Online mode}
As mentioned in Sec.~\ref{sec:user_interaction}, the toolkit allows users to make blendshape-level adjustments.
In the HITL Online mode, the differences between the auto estimated blendshape values and the adjusted values are stored as user preferences in the toolkit.
Subsequently, with a single click, users can apply these preferences to all video frames.
The algorithm behind this design is simple yet effective.
We average all the user adjusted difference values as a user preference $\delta$, then apply the $\delta$ to the corresponding blendshapes of all the remaining frames.
Therefore, a few-to-many functionality is achieved, \textit{i.e,} you only tune few frames to inject your preference to the the whole animation.
The online mode significantly reduces the time and cost for animators to adjust animations, thereby reducing a substantial portion of their workload.

\paragraph{Offline mode}
After several rounds of online HITL interactions, the toolkit accumulates a sufficient amount of human feedback and adjustments for the blendshapes.
These data can be regarded as valuable human annotation and be further exploit in the HITL Offline mode.
Specifically, we utilize the annotation data as supervision signal to finetune the adapter model.
Considering that this data contains enough human preference, the resulted model not only achieves better blendshape estimation accuracy, but also improves the overall alignment to human preference.

There are some differences between the two modes.
In the Online mode, users' feedback is specifically provided for the current animation.
The adjustment can be made with relatively less time and will be displayed in real-time on the toolkit interface

In contrast, the Offline mode involves finetuning the model with the data collected from different videos.
Therefore, compared to preference for specific animation, this data embeds more generic and complete personal intention to all the animations.

From another perspective, the two modes can benefit each other.
On the one hand, the Online mode helps accumulate human feedback data for the Offline mode.
On the other hand, the offline finetuned model will be used to predict blendshape coefficients as initialization for the the online mode.
This essentially builds a data flywheel effect with human-in-the-loop.
It allows for iterative refinement by incorporating human expertise and continuously improving the performance of the system.
Consequently, the better the system is, the less human intervention is required.

\subsection{User Interface}

Figure \ref{fig:user_interface_design} illustrates the user interface of the Toolkit.
It consists of four areas, namely the input video area, virtual human area, user interaction area, and frame diagram area.

The input video area, located at the left side of the scene in a rectangular region, is capable of displaying the current video frame.

The virtual human area, situated in the central region of the interface, showcases the facial expressions and head poses of a virtual human.
This area is designed to show the automatically generated facial animation or the users' adjustment in real time.

The frame diagram area is located at the bottom of the interface within a dark gray box.
It visualizes all animation frames with a scatter plot.
The horizontal axis denotes the index number of frames, while the vertical axis represents the average value of all blendshape coefficients at the respective frame.
By default, each frame is denoted as a white dot.
Keyframes are specially represented by green dots.
While red dots represents frames that have been manually adjusted by the user and recorded as preferences by the toolkit.
The red vertical line serves as a progress bar, corresponding to the frame that displayed in the virtual human area.

The user interaction area, positioned on the right side of the interface, comprises a set of buttons designed to offer various functionalities for animation generation, visualization, playback, adjustment, and result exportation. In the subsequent sections, we elaborate on the specifics of each button and field.

\textit{``Initialize"} button.
It executes the initialization procedure of the toolkit, invoking both the base model and the adapter model to predict blendshape coefficients from input images or videos.

The $\leftarrow \rightarrow$ button.
It implements the function of scrolling back and forth between animation (and input video) frames.

The $<< >>$ button.
Play all video frames and animation frames continuously in either forward or backward direction.

\textit{``Play Anim"} button.
By clicking on this button, the toolkit will display the generated facial animation as we described in Sec.~\ref{sec:tool_auto_gene}.

\textit{``Target"} input field.
Within the designated region, users can input the index corresponding to the target blendshape they wish to modify in the current frame. For example, index 1 corresponds to the left eyelid, index 2 corresponds to the right eyelid, and so forth.

\textit{``Value"} input field.
Within the defined region, users are able to input the specific value of the blendshape coefficient they intend to modify, falling within the specified range of 0 to 1. The input value should be in the form of a floating-point number.

\textit{``Adjust Blendshape"} button.
Upon entering data in the designated target and value fields, clicking this button will save the adjusted value in the toolkit, registering it as the user's preference.

\textit{``Apply Preference"} button.
After finetuning a representative selection of frames and blendshapes, activating this button will compute user preference and extend it to all frames within the video.

\textit{``Clear Preference"} button.
This function erases user preferences that are not intended to be applied across all video frames.

\textit{``Add Keyframe"} button.
Users can manually insert keyframes that they wish to emphasize through the utilization of this button.

\textit{``Export Results"} button.
By clicking this button, the user feedback will be exported to a \textit{.json} file, which refers to the adjusted blendshapes coefficients.

\section{Evaluation}

\subsection{Evaluation of the Model}
\label{sec:eval_method}
The deep learning model, which aims to automatically generate animation from image or video input, was assessed from two aspects: the quantitative evaluation of modeling accuracy and the qualitative results of facial animation. In this part, virtual characters with different texture appearances and blendshape topologies were tested.

\subsubsection{Quantitative Evaluation of Modeling Accuracy}
To evaluate the modeling accuracy of the proposed deep learning method, we report the blendshape estimation accuracy of several design choices of our method against the ground truth blendshape coefficients.
In detail, we have several different settings of the implementation of the adapter model.
We compute the Mean Absolute Error (MAE) between the results and the ground truth, instead of using MSE as in the training process.
\begin{equation}
    \text{MAE} = ||\alpha'-\alpha||_1.
\end{equation}
This is because MSE helps stabilize the training process while MAE gives a more straightforward sense of error.

\begin{table}[t!]
\footnotesize
\caption{Experiments on different settings of the adapter model. We use Mean Absolute Error (MAE) for evaluation and comparison.}
\vspace{-3pt}
	\centering
	\begin{tabular}{ccc}
 	\toprule
	Layers & Hidden-dim & MAE \\ 
    \hline
    Linear$\rightarrow$ReLU$\rightarrow$Linear & 256 & 0.09 \\
    Linear$\rightarrow$ReLU$\rightarrow$Linear & 100 & 0.10 \\
    Linear$\rightarrow$ReLU$\rightarrow$Linear & 384 & 0.09 \\
    Linear$\rightarrow$LeakyReLU$\rightarrow$Linear & 256 & 0.09 \\
    Linear$\rightarrow$ReLU$\rightarrow$Linear$\rightarrow$Clamp & 256 & 0.08 \\
    Linear$\rightarrow$LeakyReLU$\rightarrow$Linear$\rightarrow$Clamp & 256 & \textbf{0.07} \\
    \bottomrule
	\end{tabular}
\label{tab:exp_translator}
\vspace{-12pt}
\end{table}

As shown in Table \ref{tab:exp_translator}, the first three experiments have compared the effect of hidden dimension between the two linear layers. With 256 as the baseline, results show that a smaller dimension may limit the model's performance, while no significant changes are observed when increasing the hidden-dim to 384. Then, we have explored different network structures, including replacing $ReLU$ with $LeakyReLU$, appending a $Clamp$ operator at the end, and the combination of the two operations. We can see from the table that truncating the output by the $Clamp$ operator has benefits in reducing the estimation error. The least error is achieved by using $LeakyReLU$ and $Clamp$ simultaneously.
All the settings contain a similar amount of trainable parameters, \textit{i.e,} two fully connected layers. The adapter model are designed to be lightweight for easier adaptation.

\subsubsection{Qualitative Results of Facial Animation}
To assess the facial animation quality of the deep learning method, we conduct evaluation from the following three perspectives.

- For one specific virtual character, we take various types of input to generate facial expressions.

- We test virtual characters with the same blendshape topology but different texture appearances.

- We test virtual characters with the same texture appearance but different blendshape topologies.

\begin{figure}
    \centering
    \includegraphics[width=0.85\linewidth]{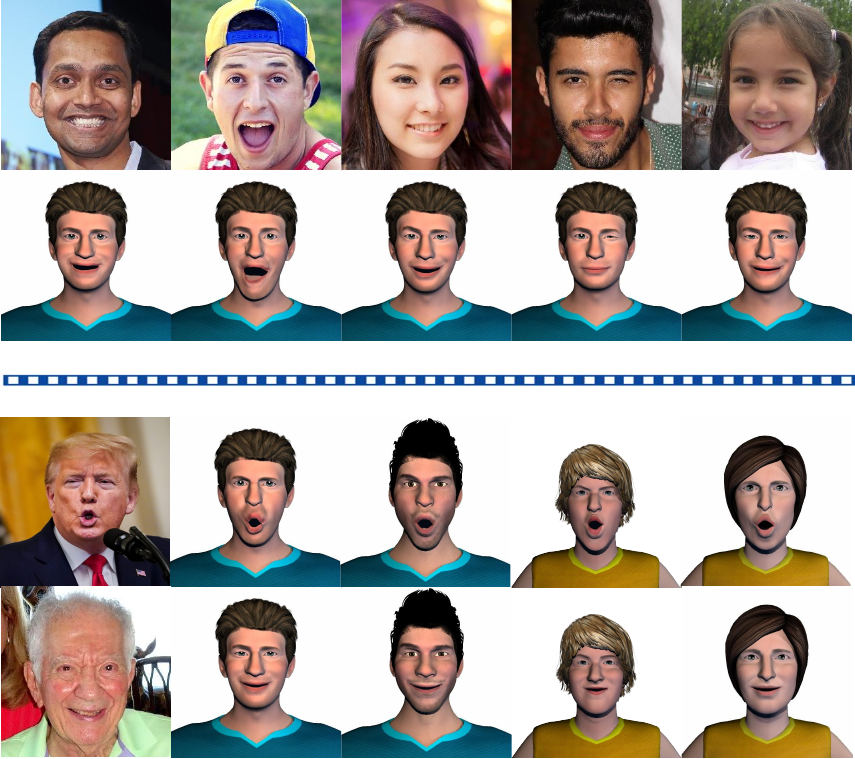}
    \vspace{-7pt}
    \caption{\textbf{Upper part}: facial animation examples of one specific virtual character.
    \textbf{Lower part}: facial animation examples of virtual characters with different texture appearances but the same blendshape topology.
    Results generated by the deep learning model fully automatically, without head pose, without human-in-the-loop intervention.}
    \label{fig:qualitative_eval_texture_ab}
    \vspace{-12pt}
\end{figure}

\begin{figure}
    \centering
    \includegraphics[width=0.85\linewidth]{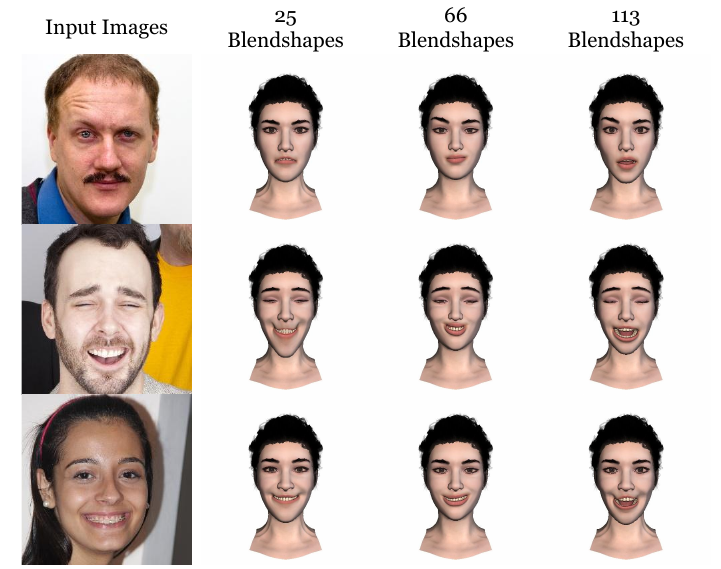}
    \vspace{-5pt}
    \caption{Facial animation examples of virtual characters with the same texture appearance but different blendshape topologies. The models are equipped with 25, 66, and 113 blendshapes, respectively. The results are generated by the deep learning model fully automatically, without head pose, without human-in-the-loop intervention.}
    \label{fig:qualitative_eval_topology}
    \vspace{-12pt}
\end{figure}

The qualitative results of animation are shown in Fig.~\ref{fig:qualitative_eval_texture_ab} and Fig.~\ref{fig:qualitative_eval_topology}.
The deep learning model takes the human face image as input and estimates a set of blendshape coefficients, which are used to render virtual human facial expression.
For one specific character equipped with 50 blendshapes, as shown in the upper part of Fig.~\ref{fig:qualitative_eval_texture_ab}, we observe that the generated faces are capable of replicating the reference facial expressions and are robust to various types of input.
For different characters equipped with 50 blendshapes, as shown in the lower part of Fig.~\ref{fig:qualitative_eval_texture_ab}, the model can be directly applied to a wide range of characters with the same blendshape topology, even with different texture appearances and identities.
One important advantage of the proposed model is its flexibility in transferring to virtual characters with various blendshape topologies.
Fig.~\ref{fig:qualitative_eval_topology} shows that our model can be transferred to characters with different topologies.
The three models are equipped with 25, 66, and 113 blendshapes, respectively.

\subsubsection{Discussion on the Effect of Blendshape Topology}
\label{sec:trade-off}
It can be observed from Fig.~\ref{fig:qualitative_eval_topology} that the expressiveness of the generated result is affected by the number of blendshapes.
In the first row of Fig.~\ref{fig:qualitative_eval_topology}, the right eyebrow of the person in the input image is raised, but this expression is not captured by the 25-blendshape virtual character. However, both the 66-blendshape and 113-blendshape virtual characters depict the raised right eyebrow. This disparity arises from the fact that the 25-blendshape virtual character's topology does not support individual eyebrow movement. 
In the second row of Fig.~\ref{fig:qualitative_eval_topology}, the person in the input image has an open lower jaw, a movement accurately represented by the 113-blendshape virtual character. However, the 25-blendshape and 66-blendshape virtual characters cannot depict this movement due to their topology lacking the necessary motion units.
In the third row of Fig.~\ref{fig:qualitative_eval_topology}, the person in the input image is displaying a smiling expression. Blendshapes representing this smile are present in all three virtual characters with different topologies. It can be observed that the 25-blendshape virtual character exhibits the most convincing smile, while the 113-blendshape character displays the least convincing expression.

In Table~\ref{tab:blendshape_topology}, we report the quantitative results of models that are adapted to different virtual characters.
In concept, the greater the number of blendshapes, the more detailed the virtual human model will be, resulting higher quality animation.
However, in Table~\ref{tab:blendshape_topology} we observe that greater number of blendshape leads to higher error, since it makes the optimization more difficult.
Actually this phenomenon also holds true for humans, \textit{i.e,} it is also more difficult for human animators to manipulate a large number of blendshapes.

The results suggest that there is a trade-off when determining the number of blendshapes of virtual characters. On the one hand, fewer blendshapes make adapter training easier, resulting in lower inference error and more precise control. However, fewer blendshapes implies that the upper bound expressiveness of the virtual character is also limited. On the other hand, more blendshapes increase the potential expressiveness of the virtual character, while also posing challenges in optimizing the adapter model.

In terms of adaptation time, with the support of NVIDIA GeForce RTX 3060 GPU hardware, training for adapting 3D virtual characters with different topological structures takes only about 30 minutes. Compared to human animators manually adjusting blendshapes to fit virtual characters with different topological structures, this method eliminates the need to consider the order, physical meanings, and controlled facial units of blendshapes in the virtual characters. This significantly reduces the time and manpower costs associated with animation editing.

\begin{table}[t!]
\footnotesize
\caption{Comparison on MAE of Various Blendshape Topologies.}
\vspace{-3pt}
	\centering
	\begin{tabular}{ccc}
 	\toprule
	MLP Architecture & The Number of Blendshapes & MAE \\ 
    \hline
    (64, 256) $\rightarrow$ (256, 25) & 25 & 0.0805 \\
    (64, 256) $\rightarrow$ (256, 66) & 66 & 0.1063 \\
    (64, 256) $\rightarrow$ (256, 113) & 113 & 0.1076 \\    
    \bottomrule
	\end{tabular}
\label{tab:blendshape_topology}
\vspace{-10pt}
\end{table}

\subsection{Evaluation of the Toolkit}
\label{sec:eval_tookit}
For the toolkit, different HITL(human-in-the-loop) modes for adjusting animations were evaluated on their functional usability first and then a pilot study was conducted to gather opinions and suggestions from some end-users regarding the using of the toolkit.

\subsubsection{Usability of different HITL modes}

\begin{figure}
    \centering
    \includegraphics[width=0.95\linewidth]{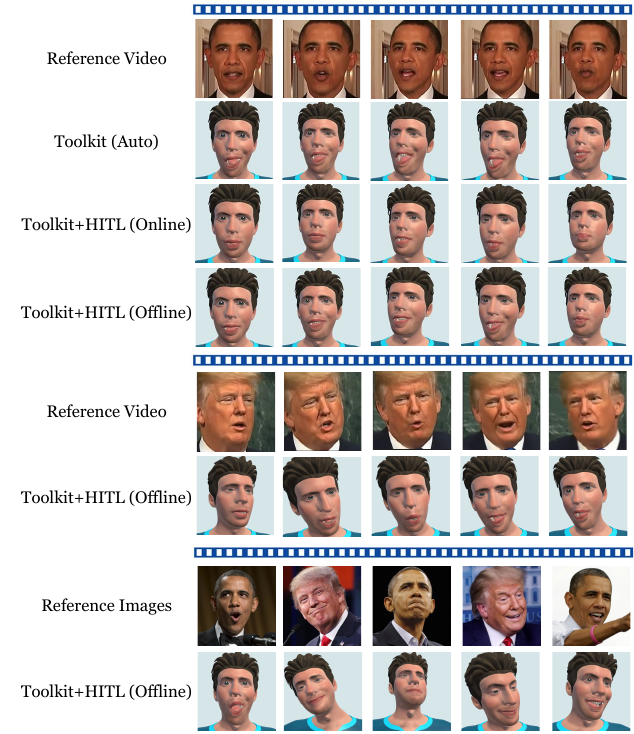}
    \vspace{-5pt}
    \caption{Facial expression animations generated by the toolkit. The virtual character is able to replicate the the facial expression and head pose of the input image or video.}
    \label{fig:examples_toolkit}
    \vspace{-14pt}
\end{figure}

In this part, we use the virtual character with 50 blendshapes.
Both facial animation quality and satisfaction score under three different modes (Auto, HITL Online, HITL Offline) were compared.

\paragraph{Facial animation quality}
We provide the qualitative results generated by the toolkit in Fig.~\ref{fig:examples_toolkit}.
In the first group, we qualitatively compare the results of three different modes mentioned above.
It can be observed that the animation results generated by HITL are closer to the reference video than the ones auto-generated by the toolkit.
However,the differences between the two modes of the HITL are subtle, which indicates the model can be offline tuned into similar performance with the online tuning one.
In the below part of Fig.~\ref{fig:examples_toolkit}, we provide more examples generated by the toolkit with various input identities and formats, including videos and images.
These examples demonstrate that the toolkit is capable of replicating the facial expression and head pose of the input.

\paragraph{Satisfaction score}
We also compare human satisfaction scores about the quality of virtual human animations generated by the toolkit under three different modes:
(1) Auto. (2) HITL Online. (3) HITL Offline.
13 participants (8 males, 5 females), which were college students aged 18 to 25 years old with various majors, were asked to give a satisfaction score ($1-10$) to the facial animation images.
Specifically, each participant was required to complete six sets of rating tasks. In each set of rating task, the participants were shown one real human facial image and three different animation images generated in three different modes according the real human image. They were then asked to rate each animation image based on how visually reasonable and close it appeared to the real image. A rating scale of 1 to 10 was used, where a score of 1 indicated that the animation image was completely dissimilar to the real image, while a score of 10 indicated that the animation image closely replicated the real image. It is important to note that participants were unaware of which animation image was generated using which mode, and the placement position of the three different animation images was randomized in each setting of rating task.

The satisfaction score for a particular mode of the toolkit was calculated by averaging the six sets of scores given by all the participants. As shown in Table~\ref{tab:main_table}, the satisfaction score is highest in the Toolkit (HITL Online) mode and lowest in the Toolkit (Auto) mode. This indicates a significant improvement in the animated effects after incorporating the human-in-the-loop (HITL) mode, as it integrates human preferences, aligning better with users' cognition and satisfaction levels compared to the original animations. The satisfaction scores in the Toolkit (HITL Offline) mode are very close to those in the Toolkit (HITL Online) mode.
This suggests that the mode that involves human feedback into the model training process offline, namely the Toolkit (HITL Offline), generates animated effects comparable to those generated in the Toolkit (HITL Online).

\paragraph{Generation time}
In terms of generation time, as shown in Table~\ref{tab:main_table}, the animation generation time is shortest in Toolkit (Auto) mode, while the generation time in Toolkit (HITL Online) mode is approximately four times longer than that in Toolkit (Auto) mode. Although this represents a significant speed improvement compared to the manual editing of animations by animators, the animation generation time in Toolkit (HITL Online) mode is still relatively long. In Toolkit (HITL Offline) mode, on the other hand, only a single, relatively long training period is required. After this training time, animations can be generated at the speed of Toolkit (Auto) mode, and the quality of the animations is relatively good.

In summary, the Toolkit (HITL Online) mode and Toolkit (HITL Offline) mode has their own unique advantages and complement each other.
Toolkit (HITL Online) shows the best satisfaction score, with the cost of longer generation time.
While Toolkit (HITL Offline) achieves satisfaction score that is very close to the Online mode in a shorter time.
Toolkit (HITL Online) provides valuable human annotation to tune the model of Toolkit (HITL Offline), while Toolkit (HITL Offline) can reduce the tim required for manual adjustments in Toolkit (HITL Online) thanks to the improved performance.

\begin{table}[t!]
\footnotesize
\caption{Comparison on frame-wise animation generation time cost and satisfaction score.}
\vspace{-3pt}
	\centering
	\begin{tabular}{ccc}
 	\toprule
	Mode & Score & Generation Time \\ 
    \hline
    Toolkit (Auto) & $4.47\pm1.93$ & $0.21\pm0.02s$  \\
    Toolkit (HITL Online) & $7.07\pm1.95$ & $0.84\pm0.12s$  \\
    Toolkit (HITL Offline) & $6.57\pm2.27$ & Training +  $0.21\pm0.02s$\\
    \bottomrule
	\end{tabular}
\label{tab:main_table}
\vspace{-12pt}
\end{table}

\subsubsection{Experience of using the toolkit}
Finally, we conducted an independent pilot user study where we invited a group of end-users to try out the proposed toolkit. The purpose was to gather their opinions and suggestions regarding the toolkit's usage to identify any areas for improvement.

\paragraph{Participants}
16 participants (6 male, 9 female and 1 non-binary gender; age $M$=24.6, $SD$=3.12) were recruited and completed the user study. All of them are college students who have experience and familiarity with 3D software applications such as MAYA, 3DMAX, and Unity3D. This choice was made considering that they are potential end-users of the toolkit and their experience with similar software may enable them to provide valuable and constructive opinions and suggestions for the improvement of the toolkit.

\paragraph{Procedure}
Before starting, each participant was given an introduction about the study and was required to sign a consent form. Following this, they were asked to provide the demographic information (age, gender, occupation, 3D software experience).
After that, they were instructed to watch an instructional video (6 minutes) that provided an overview of the toolkit's functions and user interface. 
This step ensured that all participants had a basic understanding of how to use the toolkit before proceeding further, similar to the approach taken with commercial tools or systems.

Subsequently, participants were asked to first utilize the toolkit (Auto) and then try the toolkit (HITL online) to give manipulations.
When using the toolkit, participants were given the freedom to either follow the examples (4 example videos and 19 images) provided in the instructional video or explore the toolkit independently based on their own preferences. 
This approach allowed participants to have flexibility in their usage and ensured that their experience with the toolkit was aligned with their individual needs and goals.
Participant was asked to try all the functions provided by the toolkit and generate animation faces by using at least one video and five images. 

After the trial using of the toolkit, participants were asked to complete : 1) a standard SUS questionnaire~\cite{brooke1996sus} measuring the usability of the toolkit ($1-5$ score), 2) a semi-structured interview to gather opinions and suggestions about the toolkit.
It requires approximately 40 to 60 minutes to complete the study and each participant was rewarded with 10 USD for their participation.
 
\paragraph{Results}
With the standard SUS questionnaire~\cite{brooke1996sus}, participants find that the toolkit is well integrated (Mean=4.0, Std=0.79), easy to use (Mean=4.4, Std=0.77) and are willing to use this toolkit in their work  (Mean=4.1, Std=0.93).
They feel that they can have a good command of the toolkit and use it confidently after watching the instructional video provided (Mean=4.6, Std=1.00).
The SUS results show that the toolkit is easy to learn and use. 

Participants' opinions and suggestions are gathered through the semi-structured interview. Following are some interview questions along with the corresponding answers.

``\texttt{How do you think about the facial animation generation? Feel free to discuss topics such as character design, animation quality, or any other aspects you find interesting.}"

``\textit{The expression animation automatically generated by this tool can simulate most of the facial expressions, especially for faces with relatively large facial expressions. The eyes and mouth need to be improved, and the micro-expressions of the face, such as the face around the nose might be a little stiff and unnatural.}"

``\textit{The general facial expression and pose retargeting looks great, but there are some flaws on detailed regions, like the eyes, mouth.}"

``\textit{Virtual expressions are closer to real videos or images and can meet basic needs. But if you want to meet more professional and refined fields such as animation production, you may need more realistic expressions.}"

``\textit{The character design mainly leans towards a cartoon style, with support for clear facial expressions. However, there is room for improvement in generating subtle facial expressions, which might require more complex and realistic virtual characters.}"

``\textit{If the toolkit could showcase animations of virtual characters' body postures and movements, it might be more easily applicable and widely adopted.}"

``\texttt{How do you think about the HITL function for human adjusting? Have you made further modification on the automatically generated animation?}"

``\textit{This toolkit can roughly replicate things in the real world. I think the expressions generated in some regions are too exaggerated, but when you can use the adjustment tool to adjust the parameters, the exaggeration is not unbearable. This is already good enough to reduce a lot of workload.}"

``\textit{I manually edited the animations generated by the toolkit because I preferred my virtual character to have a friendly appearance. So, I adjusted the facial parameters controlling the smile on the cheeks manually and tweaked only a few parameters. Then, I could quickly deploy the smiling expression across all frames using the ``Apply Preference" feature. This method significantly reduced my workload compared to adjusting the smile expression for each frame individually. However, its precision still needs improvement.}"

``\textit{I prefer my virtual character's animations to be more precise, so I manually adjusted the eyes and eyebrows of the character to keep them in line with the positions in the video.}"

``\texttt{What are the advantages and disadvantages of this toolkit? Please feel free to provide any suggestions for improving the toolkit.}"

The advantages of the toolkit that participants acknowledged can be summarized as follows:
1) It's easy and also quick to generate an acceptable facial animation based on an image or video input.
2) It is favorable to allow users to adjust the generation with which they are not satisfied and also make modifications to align with their preferences.
3) The tool can showcase virtual human expressions from any angle, a capability that 2D virtual characters cannot achieve.
4) The tool does not have any specific requirements for input face data, making it applicable to various characters, which may serve as a bridge between 2D videos and 3D models.

The disadvantages that participants mentioned can be summarized as follows:
1) The facial animation details are not always accurate in certain situations. For instance, there may be slight discrepancy in the positioning of the eyes and eyebrows compared to their real positions.
2) The tool supports characters with relatively limited facial detail and the button layout of the interface is not intuitive enough.
3) The tool is currently limited to use in Unity and cannot be utilized in software such as Unreal Engine or MAYA.

The suggestions that participants given can be summarized as follows:
1) The current facial animation quality is acceptable, but there is still room for improvement, particularly in terms of enhancing facial details.
2) It would be beneficial to support virtual characters with both facial expressions and body movements, as this would enable a broader range of applications.
3) It is recommended to generate facial animations using more natural stimuli, such as texts, audios and facial images or videos with artifacts.
4) The toolkit is also expected to be deployable on various platforms.

\section{Limitation}
Our work has the following limitations.
Firstly, there is still a large space for improvement on the quality of generated animations, especially when comparing to commercial softwares.
For example, the toolkit struggles to replicating very detailed facial expressions.
This approach does not aim to completely replace human animators or compete with commercial software, but to assist human animators when operating their personalized virtual human characters. In production, we still need human animators to supervise and improve the quality of the animation.
And in future work, we will optimize our algorithms and toolkit to enhance the accuracy and inference speed of the animations.

Secondly, in the evaluation of our method, we mainly focus on most common humanoid characters and assume that a clear input is given.
This can cover a wide range of application scenarios.
Investigation on diverse characters would be a future work, including characters with different genders, skin tones, facial artifacts (like jewelry, tattoos, scars, blemishes), and non-human characters.

Thirdly, the evaluation section does not include comparisons with other methods. The main reason is that related works are either close-sourced software or designed for fixed virtual characters, making an apple-to-apple comparison prohibitive. Our solution is not coupled with specific characters, enabling flexible customization. We hope this work serves as an initial baseline and inspires future works.

Fourthly, the user study presented in the paper is a pilot study demonstrating some properties of the proposed approach.
It provides a preliminary assessment of the the toolkit and provides some guidelines for future direction.
An improved toolkit with a more thorough user study is a future work.

\section{Conclusion}
In this paper, we propose a holistic solution for automatic facial animation generation of customized characters regardless of their blendshape topologies and texture appearances.
This is achieved by estimating the blendshape coefficients of the input image or video.
We first propose a deep learning model to estimate the blendshape coefficients of the reference facial expression in the given image.
Then we develop a toolkit that encapsulates the deep learning model with user-friendly interfaces and human-in-the-loop scheme.
The evaluation results of the deep learning method indicate that the proposed solution offers the flexibility to support customized virtual character models.
Moreover, the developed toolkit enables users to generate facial animations in an easy and efficient manner, resulting in acceptable animation quality. When involving human feedback, \textit{i.e,} human-in-the-loop, the performance of the solution can be further improved.
We make the code public to benefit the animators and inspire further study in the domain of virtual character animation.

\section*{Acknowledgement}
This work was supported by the National Key R\&D Program of China (2022ZD0117900), the National Natural Science Foundation of China (62332015, 62302494), and the Open Research Fund of Guangxi Key Lab of Human-machine Interaction and Intelligent Decision (GXHIID2201).

\clearpage
\bibliographystyle{abbrv-doi}

\bibliography{template}
\end{document}